\newcommand*{\ket}[1]{\ensuremath{|#1\rangle}}

\newcommand*{\ketup}{\ensuremath{\ket{\!\!\uparrow}}}
\newcommand*{\ketdown}{\ensuremath{\ket{\!\!\downarrow}}}

\documentclass[aip,rsi,reprint,groupedaddress,nofootinbib]{revtex4-1}
\pdfoutput=1

\usepackage{epstopdf}
\usepackage{graphicx}
\usepackage{hyperref}
\usepackage{amssymb,amsmath}

\begin{document}
\title{Superradiant Raman Laser Magnetometer}
\author{Joshua M. Weiner}
\author{Kevin C. Cox}
\author{Justin G. Bohnet}
\author{Zilong Chen}
\author{James K. Thompson}
\affiliation{JILA, NIST and Department of Physics, University of Colorado,\\ Boulder, Colorado 80309-0440, USA}

\begin{abstract}
We demonstrate a proof-of-principle magnetometer that relies on the active oscillation of a cold atom Raman laser to continuously map a field-sensitive atomic phase onto the phase of the radiated light.  We demonstrate wideband sensitivity during continuous active oscillation, as well as narrowband sensitivity in passive Ramsey-like mode with translation of the narrowband detection in frequency using spin-echo techniques. The sensor operates with a sensitivity of 190 pT/$\sqrt{\text{Hz}}$ at 1 kHz and effective sensing volume of $2\times 10^{-3}$ mm$^3$. Fundamental quantum limits on the magnetic field sensitivity of an ideal detector are also considered. 
\end{abstract}
%
%\date{\today}
%
\maketitle
Magnetometers are useful in a broad range of applications, including NMR~\cite{Greenberg98,Laloe69}, fundamental physics\cite{Ivanov99,Yoshiki10}, bio-sensing\cite{Kuhara99}, and atmospheric physics\cite{Le95}. A wide array of physical systems with various trade-offs in sensitivity, bandwidth, size, and operating conditions have been employed.  Examples include magneto-resistive materials~\cite{Kerr04}, SQUIDs~\cite{Weinstock96}, and nitrogen-vacancy centers in diamond~\cite{Lukin08}.  Atom-based sensors offer several advantages:  intrinsically high sensitivity to magnetic fields, well-determined absolute calibration and scale factors, no need for a cryogenic environment, and remote sensing capabilities via optical probing~\cite{Budker06,Budker07,Kitching04,StamperKurn11,Mikhailov12}.  

In this Letter, we demonstrate a unique atomic magnetometer based on Raman lasing transitions between hyperfine ground states of an ensemble of $10^6$ laser-cooled $^{87}$Rb atoms confined in a low-finesse optical cavity~\cite{Bohnet12a,Bohnet12b,Bohnet12c}. The lasing frequency is sensitive to magnetic fields rather than the cavity resonance frequency because the laser operates deep in the bad-cavity, or superradiant, regime where the atomic coherence damping rate $\gamma_\perp$ is much smaller than the cavity damping rate $\kappa$. 
 
The vast majority of atomic magnetometers  sense an atomic response to magnetic fields through the polarization rotation or phase shift of probe light passed through an atomic vapor~\cite{Budker07,Lukin08a,Romalis11,Polzik10}.   In contrast, here we directly determine the phase response $\phi(t)$ of an atomic dipole to external magnetic fields from the phase of the optical radiation $\psi(t)$ emitted by the dipole. 

We demonstrate that our sensor can transition between periods of active oscillation and passive Ramsey-like phase evolution, enabling the detection of magnetic fields free from possible perturbations, and with high repetition rates as was first discussed in Ref.~\onlinecite{Bohnet12b}.  This flexibility is realized by controlling the optical radiation rate via the intensity of a Raman dressing laser. Unlike typical good-cavity lasers, the passive mode of oscillation is possible in this bad-cavity laser because the atomic gain medium is the primary reservoir of phase information. 

In the following, we will also consider two fundamental limitations on the magnetic field sensitivity of an ideal superradiant magnetometer: quantum Schawlow-Townes phase diffusion of the atomic phase $\phi(t)$ at low frequencies, and photon shot noise (PSN)  associated with detecting the emitted light phase $\psi(t)$ at higher frequencies\cite{Bohnet12a,Bohnet12b}.  We demonstrate that operation in the passive mode, combined with spin echo techniques, allows the enhanced sensitivity of the low frequency detection band to be translated to higher frequencies. In principle, operation in the passive mode also avoids noise limits set by fundamental quantum phase diffusion, and may theoretically achieve sensitivity at the standard quantum limit on phase estimation using coherent spin states\cite{Bohnet12b}.
\begin{figure}
\includegraphics[width=3.375in]{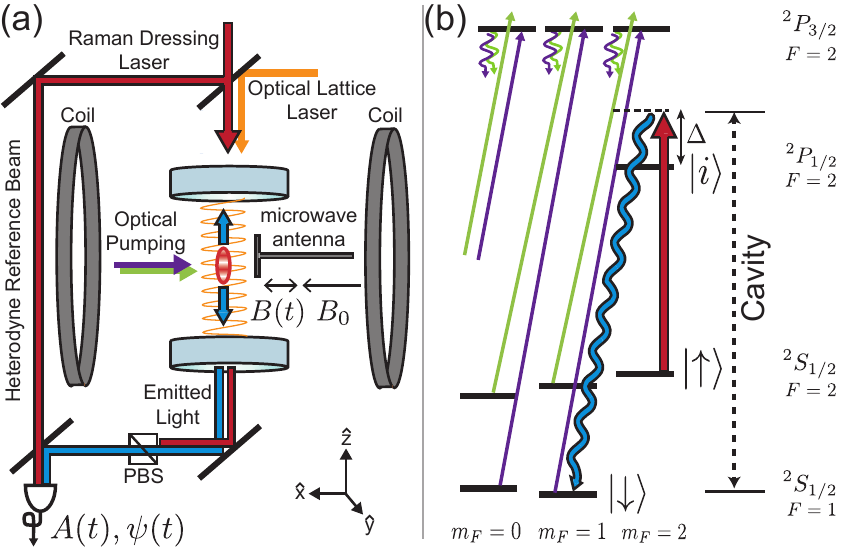}
\caption{\label{fig:fig1}(a) Experimental schematic and (b) energy level diagram.  The laser-cooled atoms are trapped between the cavity mirrors, and the coils are used to apply bias and test magnetic fields. Atoms are continuously optically pumped back to $\ketup$~using two lasers applied from the side (green, violet).  The dressing laser (red) induces a Raman decay to $\ketdown$, with collectively enhanced emission into the cavity mode (blue).  Both the trapping optical lattice and dressing beams are filtered from the cavity output before the light is detected at the photodiode using a heterodyne reference beam derived from the same laser as the dressing laser.}
\end{figure}

The superradiant magnetometer described here operates in a similar manner to the bad-cavity laser previously used to explore physics~\cite{Bohnet12a,Bohnet12b,Bohnet12c} underlying proposed $\sim 1$ mHz linewidth lasers \cite{CHE09,Meiser09,Meiser10}. A simplified experimental diagram and energy level diagram is shown in Fig.~\ref{fig:fig1}. The atomic gain medium of the laser consists of $N = 1$ to $2\times10^6$ $^{87}$Rb atoms laser cooled to 40~$\mu$K  and trapped in a one-dimensional optical lattice at wavelength 823 nm. The laser cavity has a finesse $F = 710$ with mirror separation $L=1.9$~cm and TEM$_{00}$ mode waist $w_\circ=71~\mu$m.  The root-mean-squared (rms) extent of the cloud is approximately $1.5$ mm along the cavity axis and $15~\mu$m perpendicular to the cavity axis, yielding an effective field sensing volume of $2.1\times10^{-3}$ mm$^3$. 

Previous studies of steady state superradiance\cite{Bohnet12a,Bohnet12b,Bohnet12c} have utilized the pseudospin-1/2 system composed of the magnetic-field-insensitive clock states. Here, we use the first order magnetic field-sensitive states $\ketup \equiv \ket{5^2S_{1/2},F=2,m_F=2}$ and $\ketdown \equiv \ket{5^2S_{1/2},F=1,m_F=1}$ to form the pseudospin-1/2 system.  The atomic dipole of this two-level system only very weakly radiates at the microwave hyperfine frequency difference between the two states $\omega_{\text{hf}}/2\pi=6.834$~GHz.  To induce an effective optical decay from $\ketup$ to $\ketdown$ at a controllable single-particle rate $\gamma$, we apply a $\pi$-polarized Raman dressing laser beam that is non-resonant with the optical cavity and detuned $\Delta/2\pi = 1.1$ GHz from the $\ketup$ to $\ket{i}\equiv\ket{5^2P_{1/2},F'=2, m_F'=2}$ transition with wavelength 795~nm. The cooperativity parameter of cavity QED, or Purcell factor, is $C = 7.7\times10^{-3}$ for the $\ketup$ to $\ketdown$ Raman transition. 

A TEM$_{00}$ mode of the optical cavity is tuned within 5 MHz of resonance with the emitted Raman photon frequency.  Single particle scattering into the cavity is too weak for useful sensitivity, with a total scattering rate from the ensemble scaling as $N C\gamma $.  However, when the total single particle scattering rate into the cavity exceeds the atomic coherence dephasing rate $\gamma_\perp$, the atoms spontaneously synchronize their dipole moments, leading to a stimulated enhancement of the emission rate scaling as $N^2 C\gamma$ as in the phenomenon of superradiance~\cite{Dicke54}.  The phase of the collectively emitted light $\psi(t)$ adiabatically follows the collective dipole moment of the ensemble $\phi(t)$ within the bandwidth $\kappa/2$, where $\kappa = 2 \pi \times 11.1$~MHz is the cavity power decay rate.  Given the resonance condition and dipole selection rules, the only available state for decay through emission into the cavity mode is the $\ketdown$ state such that there is no gain competition between different decay paths. 

To operate in a quasi-continuous mode, the atoms are continuously recycled at single-particle rate $W$ from $\ketdown$ to $\ketup$ by applying two $\sigma^+$-polarized optical repumping beams at 780 nm such that the only dark state is $\ketup$, as shown in Fig.~\ref{fig:fig1}b.

The phase noise of the 60~kHz FWHM linewidth Raman dressing laser is removed from the signal by detecting the emitted light in heterodyne using a $40~\mu$W beam also derived from the dressing laser. The signal is quadrature demodulated and both demodulation channels are recorded in the data acquisition system to reconstruct the full light phasor $\mathcal{E}(t)= A(t) e^{i\psi(t)}$. 

A set of Helmholtz coils applies a dc bias magnetic field $B_0 = 2.4\times 10^{-4}$ T in the $\hat{x}$-direction perpendicular to the cavity axis.  This establishes the quantization axis and shifts the transition frequency by $\omega_{\text{dc}}/2\pi = 5.1$ MHz relative to the zero-field ground state hyperfine splitting.  The dc field also breaks the degeneracy of the $m_F$ levels so that microwave rotations can be performed exclusively on the $\ketup$ and $\ketdown$ states with characteristic Rabi frequency $\Omega/2\pi\approx 35$ kHz for the spin echo studies to follow.
\begin{figure}
\includegraphics[width=3.375in]{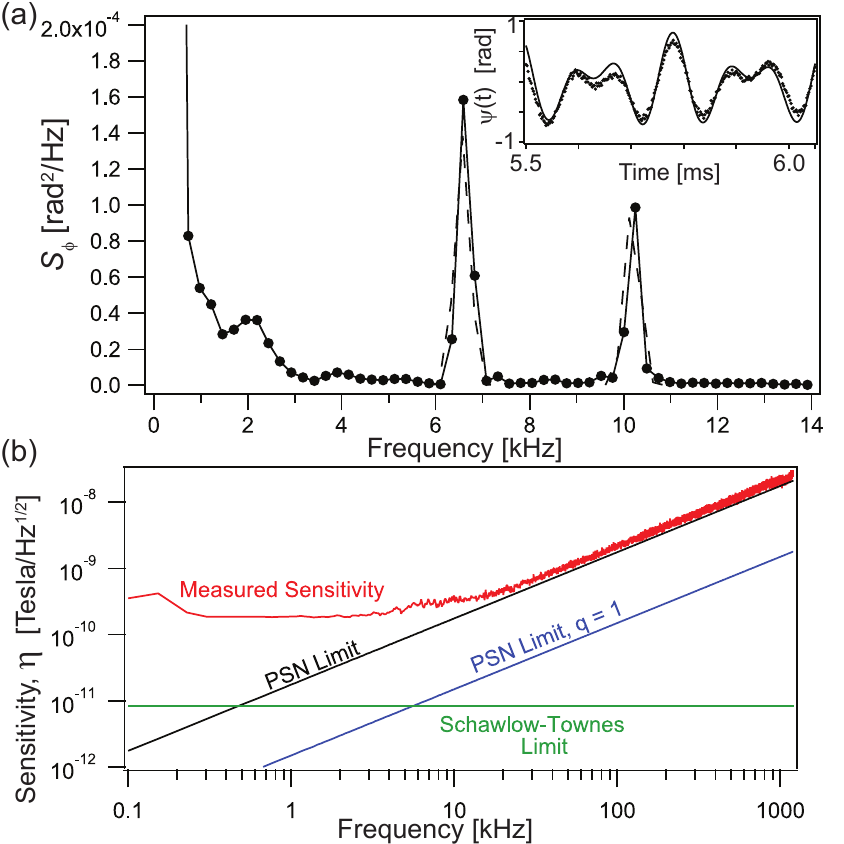}
\caption{\label{fig:fig2}(a) The power spectral density (PSD) of $\mathcal{E}''(t)$ in the presence of modulation at 6.9 and 10.2 kHz, averaged over many trials. Dashed line shows the expected PSD from the applied modulation. The inset shows the detected (points) and predicted (line) light phase $\psi(t)$ versus time. (b) The calculated magnetic field sensitivity (red) versus frequency from detection of $\psi(t)$ during active laser oscillation. Also shown are the expected Schawlow-Townes phase diffusion limit (green), the (PSN) limit (black) for $q=7.2\times 10^{-3}$, and the PSN limit for $q = 1$ (blue).}
\end{figure}

We investigate the system's response to small magnetic fields by applying small modulations $B(t)$ to the current through the same coils, with phase, frequency and amplitude set by the data acquisition computer that also records the generated light signals. The atomic transition energy $E(t)$ and hence the time rate of change of the collective atomic phase depends on the small modulation of the field through 
\begin{equation}
E(t)/\hbar = \frac{d\phi(t)}{dt} = \omega_{\text{hf}}+2\pi \alpha \left(B_0+B(t)\right)
\end{equation}
\noindent where $\alpha = 2.1\times 10^{10}$ Hz/T, and $\hbar$ is the reduced Planck constant.

We apply a small sinusoidal modulation of the field $B(t) = B_m\cos(2\pi f_m t + \theta_m)$ to create a modulation of the atomic phase as $\phi(t) =  \beta\sin(2\pi f_m t + \theta_m)$, where the modulation index is $\beta =\alpha B_m/f_m$. Here and in the rest of the paper, we ignore the contributions to $\phi(t)$ from the hyperfine splitting and the dc bias field. 

Fig.~\ref{fig:fig2} demonstrates the simultaneous detection during continuous superradiance of two applied discrete modulations at frequencies 6.9 and 10.2 kHz. Also, we observe no modulation at the difference or sum of the two modulation frequencies, confirming the linearity of this detection scheme at the current level of sensitivity.

The data was obtained by detecting the light phasor $\mathcal{E}(t)$ during a time window of 3~ms.  Atom loss causes the superradiance to cease after approximately 20 to 100~ms~\cite{Bohnet12a}, after which the atoms are reloaded and the experiment is repeated every 1 second.  The observed carrier frequency chirps as atoms are lost due to cavity pulling of the oscillation frequency combined with atom dependent shifts of the cavity mode frequency\cite{Bohnet12a}. To avoid spectral components from the chirping, we apply a third-order digital Bessel high pass filter to the recorded phase $\psi(t)$ with pass-band cutoff frequency 300~Hz to produce a filtered time phase $\psi'(t)$. To remove any possible amplitude modulation (AM) sideband noise, we set $A(t) = 1$ in the data analysis to construct $\mathcal{E}'(t) = e^{i \psi'(t)}$. This could be accomplished in real time by phase locking an oscillator with low AM noise to the emitted light phase.  Spectral leakage of the carrier due to a square sampling time window is mitigated by multiplying the phasor with a fourth-order Blackman-Harris time window $w(t)$ such that $\mathcal{E}''(t) = w(t) \mathcal{E}'(t)$.  

The double sideband power spectral density of phase fluctuations $S_\phi(f)$~\cite{Riehle04} is calculated from the Fourier transform of $\mathcal{E}''(t)$ and averaged over many trials. The spectrum of the equivalent rms magnetic field noise density, or sensitivity, $\eta(f)$ (units of Tesla/$\sqrt{\text{Hz}}$) is then calculated as 
\begin{equation}
\eta(f) = \frac{f}{\alpha}\sqrt{S_\phi(f)}.
\end{equation} 

\noindent Fig.~\ref{fig:fig2}b shows the calculated ideal detection sensitivity and the measured field sensitivity for $\gamma = 37~\text{sec}^{-1}$, quantum efficiency $q=7.2\times 10^{-3}, N = 1.1\times10^6,$ and $W = 4.5\times10^4~\text{sec}^{-1}$. In our unshielded apparatus, ambient magnetic field noise, noise in the coil driver electronics, and emission frequency chirping  contribute noise far above the Schawlow-Townes limit at frequencies $< 11$ kHz.  At high frequencies, PSN limits our ability to resolve ac magnetic fields.

In the absence of technical noise, the ideal phase noise density $S^{\text{i}}_\phi(f) = C\gamma/(\pi f)^2 + 1/(2 \dot{m}_d)$ is expected to be fundamentally limited at low frequencies by quantum phase diffusion of the collective Bloch vector (the Schawlow-Townes limit, first term) and at high frequencies by PSN (the second term).  For photon quantum detection efficiency $q$ and optimum repumping rate $W_\text{pk} = NC\gamma/2$~\cite{Bohnet12a}, the rate of detected photons is $\dot{m}_d= q R N^2 C \gamma/8$. Here, $R$ is a reduction factor for the multi-level $^{87}$Rb scheme presented with value $R \leq 3/5$. The ideal field sensitivity can be written as:
\begin{equation}
\left(\eta^{\text{i}}(f)\right)^2 = \frac{2 C \gamma}{\pi^2 \alpha^2}\left(1 + \frac{f^2}{f_\circ^2}\right)
\end{equation}
\noindent where the corner frequency is given by $f_\circ = \sqrt{q R} N C \gamma/(2 \pi)$.  The minimum detectable field scales with the single particle decay rate into the cavity $C \gamma$, while the corner frequency scales with the collectively enhanced single particle scattering rate into the cavity $N C \gamma$. In principle, the tunable decay rate $\gamma$ can be reduced until the single-particle transition broadening described by a transverse coherence decay rate $\gamma_\perp$ is no longer negligible compared to $W/2$, setting a minimum $C\gamma\sim\gamma_\perp/N$ for which $\left(\eta^{\text{i}}(f)\right)^2 \sim (\gamma_\perp/N)(1+f^2/f_\circ^2)$ with corner frequency $f_\circ\sim \gamma_\perp$. These ideal scalings are equivalent to the scaling of the standard quantum limit for unentangled atoms in the presence of transition broadening.
\begin{figure}
\includegraphics[width=3.375in]{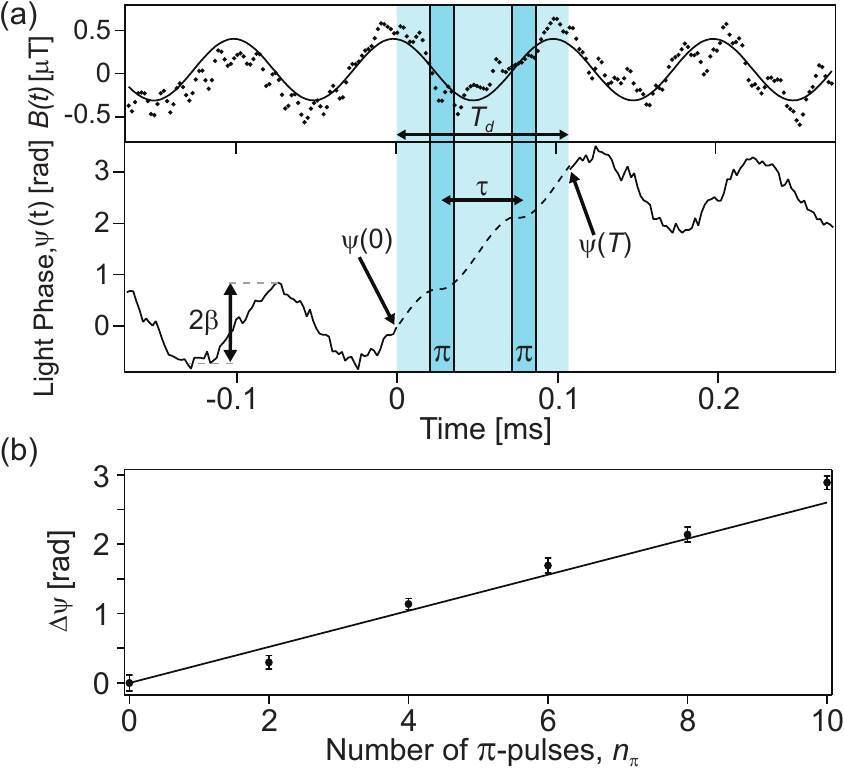}
\caption{\label{fig:fig3}(a) (Top) Time traces of the measured (points) and fitted (solid line) $B(t)$ and (bottom) the measured (solid line) light phase $\psi(t)$ and expected atomic phase $\phi(t)$ (dashed black line) in the presence of magnetic field modulation. During the dark period in which all optical pumping and dressing lasers are shut off (light blue region), spin echo $\pi$-pulses cause the unobserved atomic phase $\phi(t)$ to coherently increase in response to the applied magnetic field modulation.  This manifests as a discontinuous advance in the phase of the light when the optical pumping and dressing lasers are turned back on. (b) The magnitude of the phase advance $\Delta\psi = \psi(T_d) - \psi(0)$ (points) increases with the number of $\pi$-pulses $n_\pi$. The total dark time is $T_d = n_\pi \tau$.  Also shown is the expected slope of $2\beta$ (line).}
\end{figure}

The sensitivity of the continuous readout can be surpassed if the system operates in a narrowband detection mode based on passive evolution, analogous to Ramsey spectroscopy. As first demonstrated in Ref.~\onlinecite{Bohnet12b}, an equivalent passive Ramsey measurement technique can be realized using steady-state superradiance, interrupted for some period of time $T_d$ during which the decay rate $\gamma$ and repumping rate $W$ are set to zero. The atomic phase precesses during this dark period with no Schawlow-Townes phase diffusion and possibly lower systematic errors. Measurement of the light phase just before and after the shutoff  allows the accumulated quantum phase to be estimated as $\phi(T_d)-\phi(0)\approx\psi(T_d)-\psi(0)\equiv\Delta\psi$.  The measurement record needed to estimate the final phase $\psi(T_d)$ also serves to estimate the phase for the next iteration of the experiment, potentially allowing high repetition rate, non-destructive Ramsey-like measurements. 

Ramsey spectroscopy detection operates with a detection band centered at zero frequency.  The favorable sensitivity of dc detection can be translated to in-principle arbitrary frequency using spin echo sequences to essentially serve the role of a mixer in a lock-in amplifier~\cite{Ozeri11}. An even number $n_\pi$ of $\pi$-pulses must be applied in order to restore population inversion so that the superradiance resumes with the Bloch vector close to its initial steady-state inversion.

Here we demonstrate the translation of enhanced sensitivity near dc to higher frequencies. Fig.~\ref{fig:fig3}a shows a sequence in which the phase accumulation is coherently enhanced through a spin echo pulse sequence with the $\pi$-pulses aligned in time to nodes of the applied modulation $B(t)$. The phase difference $\Delta\psi$ is approximately 2$\beta n_\pi$, where $\beta = 0.71~\text{rad}$ and $n_\pi = 2$ for this data, and with only small corrections due to the finite $\pi$-pulse times of $15~\mu$s. For a modulation phase that maximizes the accumulated phase deviation, the sensitivity (units of Tesla/$\sqrt{\text{Hz}}$) in a 1 Hz bandwidth with no measurement dead time is 
\begin{equation}
\eta = \frac{\sigma_\psi}{4\alpha\sqrt{T_d}}\label{eq:nbsensitivity}
\end{equation}
\noindent where $\sigma_\psi$ is the rms measurement noise of the light phase difference $\Delta\psi$. Fig.~\ref{fig:fig3}b shows the phase advance $\Delta\psi$ versus the number of $\pi$-pulses applied for a $10$ kHz modulation with $\beta = 0.13~\text{rad}$ and modulation phase for which nodes of $B(t)$ are aligned to the pulse times. The fitted slope is 0.28 rad/pulse, close to the expected slope of magnitude $2\beta = 0.26$ rad/pulse. 

\begin{figure}
\includegraphics[width=3.375in]{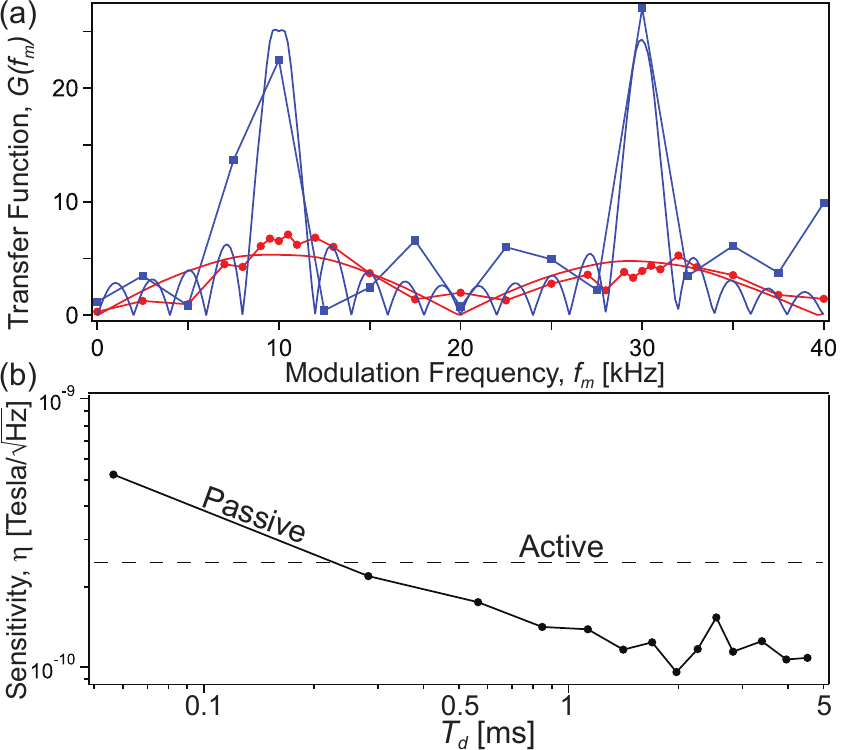}
\caption{\label{fig:fig4}(a) Normalized phase accumulation transfer functions measured for $n_\pi = \{2, 10\}$ and $\beta = \{0.25, 0.08\}$ (blue squares and red circles, respectively). (b) Field sensitivity $\eta$ versus $T_d$, calculated using Eq.~\ref{eq:nbsensitivity} and the measured detection noise $\sigma_\psi$ with average 0.45~rad. The sensitivity drops below the measured field sensitivity in the active mode (dashed line) at 36 kHz.}
\end{figure}

Fig.~\ref{fig:fig4}a illustrates both that the sensitivity can be translated in frequency, and that the sensitivity increases with the number of $\pi$-pulses, while the bandwidth is decreased.  This figure specifically shows the phase-insensitive transfer function
\begin{equation}
G(f_m)\equiv\beta^{-1}\sqrt{\left\langle\left(\Delta\psi\right)^2\right\rangle_{\theta_m}}, 
\end{equation} 
\noindent determined by measuring $\Delta\psi$ and averaging over the modulation phase $\theta_m$ at each modulation frequency $f_m$.   The spacing between the $\pi$-pulses is fixed to $\tau = T_d/n_\pi = 50~\mu$s, setting the maximum sensitivity to modulations near $10$~kHz. Two different numbers of pulses are used: $n_\pi = 2$ and 10 for the red and blue curves respectively. We find reasonable quantitative agreement between our data and modeling the effect of phase modulation by numerically integrating the Schr\"{o}dinger equation for the two-level system in the presence of a modulated classical driving field with finite Rabi frequency $\Omega/2\pi = 33$ kHz.

In Fig.~\ref{fig:fig4}b, we show that this technique allows sensitivity below the limit imposed by PSN in the active mode. We compare the field sensitivity at frequency $f = 36$ kHz using an increasing number of spin echo pulses to the PSN-limited active sensing mode at the same frequency. The phase measurement noise $\sigma_\psi$ is roughly constant as $T_d$ increases, eventually leading to a sensitivity below the PSN-limited wideband value. The difference in scaling of sensitivity shows the gain in sensitivity in the Ramsey-like detection configuration at the cost of detection bandwidth. 

In conclusion, we have demonstrated a magnetometer that utilizes the active mapping of an atomic phase onto a light field phase to operate in both an active and passive field-sensing mode, with the choice of mode dynamically selectable.  At low frequencies, the current magnetometer is limited to a minimum field sensitivity scaled to continuous operation of $190~\text{pT}/\sqrt{\text{Hz}}$.  This is approximately 30 times worse than the fundamental limit set by quantum phase diffusion, largely due to dispersive shifts of the cavity mode frequency, which in the future can be suppressed by operating at larger detunings not accessible in the current experiment, or by operation on two independent lasing modes for which the shifts are common-mode.  Future experimental work will  focus on moving to continuous operation and realizing sensitivity at the phase diffusion limit.

KC acknowledges support from NDSEG, JGB acknowledges support from NSF GRF, and ZC acknowledges support from A*STAR Singapore. This work was supported by the NSF PFC, ARO, DARPA QuASAR, and NIST. 

\end{document}